\documentclass{aa}  
\usepackage{graphicx}
\usepackage{txfonts}
\usepackage{gensymb}
\usepackage{xcolor}
\usepackage[switch, pagewise]{lineno}

\begin{document}

\title{Candidate fossil groups in the CFHTLS: a probabilistic approach
  ~\thanks{Based on observations obtained with XMM-Newton, ESO Telescopes 
    at the La Silla and Paranal Observatories, and MegaPrime/MegaCam at the
    Canada-France-Hawaii Telescope (see acknowledgements for more details).}}

\author{
C.~Adami\inst{1} \and
F.~Sarron\inst{2} \and
N.~Martinet\inst{1} \and
F.~Durret\inst{3}
}

\offprints{C. Adami \email{christophe.adami@lam.fr}}

\institute{
Aix Marseille Univ, CNRS, CNES, LAM, Marseille, France
\and
School of Physics and Astronomy, University of Nottingham, Nottingham NG7 2RD, UK
\and
Sorbonne Universit\'e, CNRS, UMR 7095, Institut d'Astrophysique de Paris, 98bis Bd
Arago, 75014, Paris, France
}

\date{Accepted . Received ; Draft printed: \today}

\authorrunning{Adami et al.}

\titlerunning{Candidate fossil groups in the CFHTLS: a probabilistic approach}

\abstract 
{Fossil groups (FGs) have been discovered 25 years ago, and are now defined
as galaxy groups with an X-ray luminosity higher than $10^{42}\
h_{50}^{-2}$~erg~s$^{-1}$ and a brightest group galaxy brighter than the other group
members by at least two magnitudes. However, the scenario of their formation remains
controversial.}
{We propose here a probabilistic analysis of FGs, extracted from the large catalog
of candidate groups and clusters previously detected in the CFHTLS survey based on photometric
redshifts to investigate their position in the cosmic web and probe their
environment.  }
{Based on spectroscopic and photometric redshifts, we estimated the probability of
galaxies to belong to a galaxy structure, and by imposing the condition that the brightest group
galaxy is at least brighter than the others by two magnitudes, we computed the probability for a given
galaxy structure to be a FG. We analyzed the mass distribution of these candidate FGs, and
estimated their distance to the filaments and nodes of the cosmic web in which they
are embedded.}
{We find that structures with masses lower than $2.4\times 10^{14}$~M$_\odot$ 
have the highest probabilities of being fossil groups (PFG). Overall,
structures with PFG$\geq$50$\%$ are located close to the cosmic web filaments (87\% are located 
closer than 1~Mpc to their nearest filament). They are preferentially four times more distant from 
their nearest node than from their nearest filament.}
{We confirm that FGs have low masses and are rare. They seem to reside closely to
cosmic filaments and do not survive in nodes. Being in a poor environment might
therefore be the driver of FG formation because the number of nearby
galaxies is not sufficient to compensate for the cannibalism of the central group galaxy.}


\maketitle

\section{Introduction}

Fossil groups (FGs) are puzzling large-scale structures which present high X-ray
luminosities but fewer bright optical galaxies than groups or clusters
of galaxies. \citet{Ponman+94} reported the first observation of such an object.
\citet{Jones+03} later defined FGs as extended X-ray sources with an X-ray
luminosity of at least $L_{\rm X}=10^{42}\ h_{50}^{-2} {\rm erg~s}^{-1}$, and a
brightest group galaxy (BGG) at least two magnitudes brighter than all other group
members. An open question is the formation of these peculiar objects and why
they present such a low amount of optically emitting matter.

An early explanation that has been proposed by \citet{Jones+03} is that FGs are
the remnants of early mergers, and that they are cool-core structures which
a long time ago accreted most of the large galaxies in their environment.
Although this scenario was supported by some hydrodynamical
simulations by \citet[e.g.,][]{D'Onghia+05}, some clues also exist that FGs might
be a temporary stage of group evolution before they capture more galaxies in their
vicinity, as reported for instance by \citet[see][based on N-body simulations]{vonBenda-Beckmann+08}.

The situation is not simpler on the observational side, partly because we lack
large samples of FGs, and partly because selection criteria differ. Fossil groups can be
studied through their X-ray \citep[e.g.,][]{Adami+18} or optical properties
\citep[e.g.,][]{Santos+07}. \citet{Girardi+14} found identical behaviors for regular
groups and FGs when they considered the relation between their X-ray and optical
luminosities, which suggests that FGs contain the same amount of optical material than
traditional groups, but that it is concentrated in a giant elliptical galaxy that
has cannibalized most surrounding bright galaxies early on. \citet{LaBarbera+09}
also found that the optical properties of BGGs in FGs are identical to those of
giant isolated field galaxies. Both analyses support the scenario that FGs are the
result of a large dynamical activity at  high redshift, but in an
environment that is too poor for them to evolve into a cluster of galaxies through the hierarchical
growth of structures. Based on Chandra X-ray observations, \citet{Bharadwaj+16}
found that FGs are mostly cool-core systems, which adds to the other indications that
these structures are now dynamically dead.

The most recent observations, however, tend to contradict these results.
\citet{Kim+18} reported that NGC 1132 is a FG with an asymmetrical disturbed X-ray
profile, and suggested that it is not dynamically passive, as expected. Similarly, \citet{LimaNeto+20}
discovered shells around the BGG of NGC~4104. Based on N-body simulations, they
showed that this FG experienced a recent merger between its BGG and another bright
galaxy with a mass of about 40\% of that of the BGG. These two examples show
exceptions to the FG nomenclature that might indicate more complex
evolutionary scenarii.

One option to better understand the evolution of these systems is to perform a
probabilistic analysis of FGs. We follow this approach here by making use of the large candidate cluster and group sample of \citet{Sarron+18}.
These authors identified a large number of structures in the 154~deg$^2$ of the Canada-France-Hawaii Telescope
Legacy Survey (CFHTLS) by applying an adaptive Gaussian filtering in photometric
redshift slices. This provided us with a large sample of groups from which it was
possible to isolate FGs in a probabilistic way instead of studying particular
objects, as is usually done in the case of FGs.

In the present analysis we also make use of the network of filaments of galaxies
defined through topological criteria after the same Gaussian filtering as obtained by
\citet{Sarron+19} to investigate the position of FGs in the cosmic web. This allows
us to probe the direct vicinity of FG candidates in observational data for the
first time.

The paper is structured as follows. In Sect. 2 we describe our data set. In Sect. 3
we derive the probabilities of the group candidates to be FGs. We measure the
properties of the FGs, in particular their positions in the cosmic web in Sect. 4,
and we conclude in Sect. 5.

Throughout the paper we adopt a $\Lambda$CDM cosmology with $\Omega_{\rm m}$ = 0.30,
$\Omega_\Lambda$ = 0.70, and H$_0$ = 70 km s$^{-1}$ Mpc$^{-1}$, and all cosmological
distances are given in comoving Mpc.

\section{Data}

\subsection{CFHTLS}

Photometric redshifts are taken from the CFHTLS T0007 data release (https://www.cfht.hawaii.edu/Science/CFHTLS/T0007/), 
which covers 154 deg$^{2}$ across four Wide fields (W1, W2, W3, and W4), observed in the
$u^{*}g'r'i'z'$ filters with the  MegaCam at the CFHT. The photo-$z$s are computed using the LePhare software \citep[]{Arnouts+99,
Ilbert+06} following the method presented in \citet{Coupon+09}. LePhare computes photo-$z$s by fitting spectral energy distributions (SEDs) to the 
five-band magnitude measurements. In the present case, 62 SED galaxy templates were used. They were obtained through 
a linear interpolation
between four templates from \citet{Coleman+80} and two starburst templates from
\citet{Kinney+96} using the VIMOS-VLT Deep Survey (VVDS) spectroscopic sample \citep[e.g.,][]{LeFevre+05}. This interpolation allows us to 
accurately sample the color-redshift space.

\indent Following the definition of \citet{Ilbert+06} and \citet{Coupon+09}, the photo-$z$ dispersion was estimated using the 
normalized median absolute deviation (NMAD) estimator,
\begin{equation}
\label{eq:1}
\sigma _{\Delta z / (1 + z_{\rm s})} = 1.48 \times \mathrm{median} \,
\left ( \frac{\left | \Delta z  \right |}{(1+z_{\rm s})} \right),\end{equation}
\noindent where $\Delta z =
z_{\rm phot} - z_{\rm s}$, where $z_{\rm phot}$ and $z_{\rm s}$  correspond to the
photometric and spectroscopic redshifts, respectively. We also define the
catastrophic failure rate $\eta$ as the percentage of objects satisfying the
criterion $\left | \Delta z  \right | \geq 0.15 \times (1+z_{\rm s})$.

\indent For our analysis, we discarded any galaxy that falls into the mask of the
CFHTLS T0007 data release. These masks correspond to areas with bright stars or
artifacts, for which magnitude measurements are inaccurate. We thus avoid including objects with poor photo-$z$ quality that would degrade our sample.

To further improve the redshift estimates of the CFHTLS Wide catalog, we correlated
it with known spectroscopic redshifts from the Six-degree Field (6dF) galaxy survey \citep{Jones+09}, 
the VVDS \citep{LeFevre+13}, the VIMOS Public Extragalactic Redshift (VIPERS) Survey \citep{Scodeggio+18}, 
the VIMOS Ultra Deep (VUDS) Survey \citep{Tasca+17}, the Galaxy And Mass Assembly (GAMA) survey
\citep{Baldry+18}, and the spectroscopic part of the Cosmic Evolution Survey (zCOSMOS) \citep{Lilly+09}. 
These data were extracted through the ASPIC public
database (\url{http://cesam.lam.fr/aspic/}).
We additionally used the NED to complement this database and extracted all available
spectroscopic redshifts in the W1, W2, W3, and W4 CFHTLS regions. We performed the
cross-correlation in RA,DEC coordinates within a 1~arcsec research box and retained the
highest confidence spectroscopic redshift in case of multiple identifications. This
led to the
addition of $\sim$ 73,000 spectroscopic redshifts to the photometric catalogs. We also used these new spectroscopic redshifts to refine the quality check of the
photo-$z$ catalog. Using Eq.~\ref{eq:1}, we estimated a statistical uncertainty of
0.07 after 3$\sigma$ clipping. We verified that the number of photo-$z$ catastrophic
failures (computed before the 3$\sigma$ clipping) is small: only 1.6$\%$ of the galaxies
present a difference between photometric and spectroscopic redshifts larger than our
3$\sigma$ clipping level.
We also know that photometric redshift precision is sometimes degraded when cluster galaxies are considered 
\citep{Guennou+10, Martinet+15}, especially in massive clusters. Considering only galaxies
within one Virial radius from the XXL-detected X-ray massive structures in the W1
field \citep{Adami+18} and within the z=[0.15;0.70] range, we still have a
precision of 0.07. This does not affect our detection
rate or purity too strongly, as was shown by \citet{Sarron+18}.

Finally, we investigated the photometric redshift precision as a function of the
parent cluster mass in order to verify that low-mass structures do not have particularly degraded photometric 
redshifts. This is not expected to be the case because the galaxy populations of these low-mass structures are 
more similar to the field populations than the ones of massive structures and are less affected by 
environmental effects. To do this, we also considered the previous XXL cluster sample and
computed the photometric redshift precision as a function of the cluster X-ray
temperature. Clusters colder than 4 keV (relatively massive structures)
exhibit a precision of $\sigma _{\Delta z / (1 + z_{\rm s})}$ = 0.069. This value decreases to 0.061 
for clusters colder than 1 keV, that is, within the group regime. Photometric redshift precision therefore 
does not affect low-mass cluster detection more than high-mass
cluster detection.

\subsection{CFHTLS AMASCFI group catalog}

The original group catalog was obtained by running the Adami, MAzure and Sarron
Cluster FInder ({\small AMASCFI}) algorithm on the CFHTLS T0007 data. Details on the
cluster and group candidate catalog and on the detection algorithm can be found 
in \citet{Sarron+18} and \citet{Adami+99}. Here, we report the salient points of the analysis and the few modifications 
we applied to the original catalog.

First, we cut the galaxy catalog into redshift slices of width typical of the photo-$z$ uncertainty and offset from 
each other by $\Delta z = 0.05$. We then applied a kernel smoothing with an adaptive smoothing scale to each slice and 
identified the peaks in
these two-dimensional density maps using Sextractor \citep{Bertin+96}. A minimal spanning tree \citep[MST, see,
e.g.,][]{Adami+99} was then applied to merge individual detections with a projected
separation smaller than 1~Mpc and a redshift difference smaller than $\Delta z =
0.06$. Finally, we provided a mass estimate (Virial mass $M_{200}$) by considering a scaling relation
between mass and richness, where the richness is defined as the number of passive galaxies brighter than $M^*+1.75$, 
where $M^*$ is the characteristic magnitude of the cluster. 
Here,  passive  galaxies are defined according to their best-fit template computed by $LePhare$
when the redshift of the galaxy is fixed at the cluster redshift \citep[see][for details]{Sarron+18}.
This scaling relation was
obtained by cross-matching our detections with the X-ray catalogs of
\citet{Gozaliasl+14} and \citet{Mirkazemi+15}. We estimated the typical $M_{200}$
uncertainty to be  $\sim 0.20-0.25 \ {\rm dex}$ in \citet{Sarron+18}.
Cluster redshifts were computed as the mean photometric redshift over individual detections linked by the MST, 
weighted by their mean galaxy densities \citep[see][for details]{Sarron+18},
and present an uncertainty of $\sigma_z = 0.025 \times (1+z)$.

A mass-richness scaling relation might bias the mass of FGs toward lower values because they exhibit 
lower galaxy counts at the bright end of the galaxy luminosity function than regular groups (see, e.g., 
\citet{Zarattini+16}). We tested this hypothesis by computing the scaling relation between the richness of each structure 
and its total luminosity by summing the individual luminosities of cluster members with the same magnitude cut as for 
the richness definition (M*+1.75). This richness-luminosity function follows a power law, and no statistical difference 
can be found between samples with different PFGs (higher than 15\% and 50\% and lower than 10\%), showing that our 
mass-richness relation is unlikely to bias the mass of the candidate FGs compared to other structures.

\citet{Sarron+18} computed the selection function for clusters with mass
$M_{200} > 10^{14} M_\odot$. The completeness and purity of this cluster candidate
sample are overall about 80$\%$ and 90$\%,$ respectively, for z$\leq$0.7. However,
these values are lower when very low mass structures are considered. Our best FG candidates (PFG$\geq$50$\%$: 15 structures, see below) are mostly at z$\leq$0.4 and have masses between 1.1$\times$10$^{13}$ and 
2.4$\times$$10^{14}$ M$_\odot$. Figure~1 of \citet{Sarron+18} rather suggests a completeness of $\leq$60$\%$ and a
purity of $\leq$80$\%$ in this redshift-mass regime.

To compute the cluster probability membership of individual galaxies, we computed a photometric redshift probability
distribution function (PDF hereafter) for each group in the AMASCFI catalog. This was made by summing the photometric redshift PDFs of 
individual galaxies that lie at a distance $d < 0.5 \ {\rm Mpc}$ (at the group redshift) of the group center.
The resulting distribution was then cut off at $z_{\rm inf} = z_{\rm min,MST}$ and $z_{\rm sup} = z_{\rm max,MST}$, the lower and upper redshifts of the 
individual two-dimensional detections linked by the MST. This was done to retain only the contribution of the group in the redshift PDF. We then computed the field contribution by summing the redshift PDFs of field galaxies and removed that contribution from our first estimates, thus forming the group photometric redshift PDF.

When multiple significant peaks separated by more than $0.05 \times (1+z)$ were found, we separated them, assigning the individual two-dimensional
detections in the MST accordingly to each peak. We chose the most prominent peak to be that of the group 
and thus that defining its redshift PDF, while the other peaks were ignored. We also changed the group position and 
redshift accordingly, considering only the individual detections in the MST corresponding to the retained peak. We note that this refining process 
was marginal and only concerned a few percent of our original detections.

\subsection{Cosmic filament catalog}\label{sec:Filcat}
When the spatial distribution of FGs was investigated relative to the
cosmic web, we used the catalog of cosmic filaments and nodes detected by \citet{Sarron+19}
in the CFHTLS.
The skeleton (filament, nodes, and saddle points) reconstruction was performed as in \citet{Laigle+18},  applying the DISPERSE 
algorithm of \citet{Sousbie11} to the two-dimensional galaxy distribution in photometric redshift slices. Details about the 
catalog we used can be found in \citet{Sarron+19}. Briefly, given the uncertainty on the photo-$z$, slices were 
chosen to be $300\ {\rm Mpc}$ thick. This is orders of magnitude larger that the typical radius of filaments ($\sim 1$ Mpc),
thus leading to projection effects in the slices that may result in false detections of filaments. An important part of the work by \citet{Sarron+19} therefore was to test the method performance.

We consider here what \citet{Sarron+19} called the 'global reconstruction', where cosmic filaments and nodes are reconstructed 
in the entire field of view of each CFHTLS Wide field. Based on mock data, they showed that in the $0.15 
\leq z < 0.7$ redshift range this reconstruction is $\sim 70\%$ complete and $\sim 90\%$ pure. We refer to their work for details on the selection function computation.

In each two-dimensional slice, DISPERSE traces the filaments of the cosmic web as a set of segments joining what it identified in the 
discrete galaxy distribution as nodes (local maximum of the distribution) and saddle points.
In the theory of structure formation, galaxy groups and clusters are expected to be found at the nodes of the cosmic web. In our 
case, considering the thickness of our two-dimensional slices, we note, however, that some apparent nodes might instead be due to the projection 
of filaments in two dimensions.

For each structure in the catalog of \citet{Sarron+18}, we thus matched its position with that of the nodes of the 
two-dimensional skeleton reconstructed at the best redshift of the structure. When a node fell within a circle of one Virial radius ($R_{200}$) of the group, it was 
considered a match and the structure was considered as indeed being at a node of the cosmic web.

\section{Probabilistic approach for detecting fossil groups}

\subsection{Estimating the structure membership probability for a given galaxy}

In contrast to \citet{Adami+18}, for example, where the cluster redshifts  were known with
a very high precision (spectroscopic redshifts),
we here relied only on photometric redshift estimates. The best we can do to estimate the
cluster membership of a galaxy is to compute the probability for the
given galaxy to be part of the cluster. We describe the two possible cases below.

(1) The galaxy only has a photometric redshift. We chose here to consider full probability distribution functions from
the official CFHTLS T0007 data release (see \url{http://cesam.lam.fr/cfhtls-zphots/index/download});

(2) The galaxy has a spectroscopic redshift. In this case, we only kept the spectroscopic redshift even when a photometric redshift was available. 
We assumed for the spectroscopic redshifts a Gaussian presence PDF with a full width at half-maximum
(FWHM) of 150 km.s$^{-1}$, typical of the redshift uncertainties of spectroscopic surveys
within the CFHTLS W1 \citep[see][]{Adami+18}.

Knowing the position on the sky, magnitude in the $r$-band, error on the magnitude and photometric 
redshift PDF for a galaxy, as well as the cluster position on the sky and the photometric redshift PDF, we implemented the method presented in 
\citet{CB16} (CB16 hereafter) to obtain the probability $P_{\rm mem}$ for the galaxy ${\rm gal}$ to belong to a group $G$. We refer to the 
original paper presenting the method for details (CB16) and provide here a brief outline of the method and of our choice of 
parameters.

In the framework developed by CB16, the probability membership is obtained using the Bayes theorem, 

\begin{eqnarray}
\begin{split}
      P_{\rm mem} \equiv P({\rm gal} \in G \ \vert \ \mathcal{P}_{\rm gal}(z), m_{\rm gal}^r,({\rm RA},{\rm Dec})_{\rm gal},
      \\
      \mathcal{P}_G(z), z_G,({\rm RA},{\rm Dec})_{G}),
\end{split}
\end{eqnarray}

\begin{equation}
    P_{\rm mem} \propto P(\mathcal{P}_{\rm gal}(z) \vert {\rm gal} \in G) \ P({\rm gal} \in G),
\end{equation}
\noindent where $P(\mathcal{P}_{\rm gal}(z) \vert {\rm gal} \in G)$ is the likelihood of observing the galaxy photometric 
redshift PDF $\mathcal{P}_{\rm gal}(z)$ knowing that the galaxy belongs to the group $G$, and $P({\rm gal} \in G)$ is the 
prior probability that the galaxy belongs to the group.

Following CB16, the likelihood is taken to be

\begin{equation}
    P(\mathcal{P}_{\rm gal}(z) \vert {\rm gal} \in G) = \int \mathcal{P}_{\rm gal}(z) \ \mathcal{P}_G(z) \ dz.\end{equation}

We have written convolutions as indefinite integrals. We note that in practice, they are discrete sums sampled at 
the redshift intervals $dz=0.02$ and taken in the range $0 < z < 6$ for which the photometric redshift PDF has been calculated.

The prior is computed using the relative number density of group and background galaxies in cylindrical shells around the group. Number densities $n$ are computed as a function of magnitude $m$ and redshift $z$ using the magnitude and photometric redshift PDFs,

\begin{equation}
    n(m,z)= \frac{1}{A(z)} \sum_{\rm gal} \mathcal{P}_{\rm gal}(z) \ \mathcal{P}_{\rm gal}(m),
\end{equation}

\noindent where $A(z)$ is the area of the shell in deg$^2$ at redshift $z$ and the magnitude and photometric redshift PDFs are sampled in bins of magnitude $dm=0.1$ and redshift $dz=0.02$.

To mitigate errors due to low number counts in small volumes, we proceeded as in CB16 and used means within a running window of $\pm 5 dm$ and $\pm \sigma_{z,95}(m,z)$, which is the median $95\%$ confidence limit on individual photo-$z$ in the CFHTLS in bins of magnitude and redshift. We refer to CB16 for more details. Following their formalism and notations, the prior probability for a galaxy ${\rm gal}$ of magnitude $m_{\rm gal}^r$ at a projected distance $r_{{\rm gal}-G}$ from the group center to be a member of group $G$ located at $z_G$ is taken to be 
\begin{equation}
        P({\rm gal} \in G) = 1 - \frac{\langle n_{\rm{bkg}}^{\rm loc}(m_{\rm gal}^r,z_G)\rangle}{\langle n_{\rm tot}(m_{\rm gal}^r,z_G,r_{{\rm gal}-G})\rangle},\end{equation}

\noindent where $n_{\rm{bkg}}^{\rm loc}$ is the background number density computed in an annulus between $3$ and $5\ {\rm Mpc}$ and $n_{\rm tot}$ is the number density in the considered shell.

The total number density as a function of 
distance to the cluster was averaged in annuli (shells) around the cluster center offset from each other by $50\ {\rm kpc}$ and with 
area of a disk of radius $r = 350\ {\rm kpc}$ at the cluster redshift. This value differs from that of CB16 ($r = 450\  
{\rm kpc}$). We chose this because the lowest-mass groups in our sample have $R_{200} \sim 350 {\rm kpc}$. This average on shells has 
the advantage of minimizing the centering error on AMASCFI clusters.\\

The final estimate of the probability membership was obtained by applying a rescaling inspired by that of CB16, that is, 
normalizing by the maximum probability that would be reached if the galaxy photometric redshift PDF $\mathcal{P}_{\rm gal}(z)$ 
were centered at the group redshift $z_G$,
\begin{equation}
    P_{\rm mem} = \frac{P(\mathcal{P}_{\rm gal}(z) \vert {\rm gal} \in G) \ P({\rm gal} \in G)}{P(\mathcal{P}_{\rm gal}(z - z_G) \vert {\rm gal} \in G)}
.\end{equation}

\subsection{Estimating the probability for a structure to be a fossil group}
\label{sec:proba}

The canonical definition of a FG can be found in \citet{Jones+03}. This
includes a two-magnitude gap between the BGG and the second brightest galaxy within
half a Virial radius, and conditions on the X-ray luminosity of the structure. We
here consider the magnitude criterion within half a Virial radius. We therefore 
computed the probability for a given structure of galaxies to be a FG. This was done knowing the previous membership probabilities and computing the probability of the 
BGG to be two magnitudes brighter than the second brightest galaxy. This calculation is complicated by the
fact that each galaxy has a certain  probability of belonging to the group. We define the following events:\\

\noindent $F$: the group is fossil,


\noindent $Y_i$: galaxy $i$ is brighter than all other group galaxies by at least two
magnitudes,

\noindent $Z_i$: galaxy $i$ belongs to the group,

\noindent $m_{ij}(2)$: galaxy $j$ has a magnitude brighter than that of galaxy $i$
by two magnitudes,

\noindent and their associated probabilities, $P(F)$, $P(Y_{i})$, $P(Z_{i})$
(computed in the previous subsection), and $P(m_{ij}(2))$. The last probability is 
easily computed by comparing the magnitudes of galaxies $i$ and $j$,

\begin{equation}
P(m_{ij}(2))=\left\{\begin{array}{rl}
0 &\mbox{if $m_i+2<m_j$} \\
1 &\mbox{if $m_i+2>m_j$}
\end{array}\right.
.\end{equation}

The probability $P(Y_i)$ corresponds to the probability that galaxy $i$ belongs to
the group \textit{\textup{and}} that every galaxy $j$ satisfying $m_i+2>m_j$ does 
not belong to the same group,

\begin{equation}
P(Y_i) = P(Z_i) \times \prod_{j} \left(1-P(Z_j)P(m_{ij}(2))\right)
.\end{equation}

We note that a galaxy $i$ verifying $Y_i$ is also the BGG because the magnitude
criterion of the FG definition is more restrictive than that of 
being the BGG (by two magnitudes).

The probability of the group to be fossil is the probability that \textit{\textup{at least
one}} of the galaxies $i$ is brighter than all other group galaxies by 
at least two magnitudes, or equivalently, that \textit{\textup{at least one}} of the $Y_i$ is true,

\begin{equation}
P(F) = \sum_{i} P(Y_i)
.\end{equation}

\subsection{Is our estimated spatial density of fossil groups realistic?}

\citet{Jones+03} found about 3 to 8$\times$10$^{-7}$ fossil groups per
Mpc$^3$ (with h$_{70}$) for the most massive groups.
The precision on the FG density in our analysis is affected by several
caveats. The mass estimates of \citet{Sarron+18} come from a scaling relation 
between richness and X-ray luminosity and therefore have a significant uncertainty
that is due to the calibration. This caveat is reasonable, however, because we only applied a cut 
in mass to select the fossil group sample according to \citet{Jones+03}. Second,
given our probabilistic approach, we cannot be certain that a given FG 
candidate is actually a fossil group. The average sample of high-probability FGs should nonetheless be representative of the true FG sample.

We therefore made an attempt to compute the spatial density of our FG.
Linearly extrapolating the Figure~1 of \citet{Sarron+18} down to low masses, we 
should have a $\leq$60$\%$ detection rate for typical fossil groups (and a purity
$\leq$80$\%$). When we sum all the percentages of being a fossil group that we
computed for all the structures within the catalog of \citet{Sarron+18}, this gives the statistical
number of real FGs in this catalog. This computation gives 28.9 FGs within the catalog. Assuming a $\leq$60$\%$ detection rate and a purity of $\leq$80$\%$, we
therefore predict about 38.6 fossil groups within the CFHTLS Wide survey.
The CFHTLS Wide sampled volume is about 2.43$\times$10$^8$ Mpc$^3$,
leading to a density of $\sim$1.6$\times$10$^{-7}$ FGs per Mpc$^3$. This is of the same order as the estimates from \citet{Jones+03}.

\begin{table}[t!]
\caption{\label{tab:liste}Number of structures for different levels of probability to be a fossil group,
assuming or not assuming a 20$\%$ contribution by the dominant galaxy.}
\begin{tabular}{crr}
\hline
BCG/BGG contribution &  None    &  20$\%$  \\ 
\hline                                          
PFG$\geq$50$\%$      & 15   & 25    \\ 
\hline                                          
PFG$\geq$15$\%$      & 62   & 85    \\ 
\hline                                          
PFG$\leq$10$\%$      & 2535 & 2510  \\ 
\hline
\end{tabular}
\end{table}

\subsection{Caveats}

A possible source of uncertainties in our study is related to the structure of
the Brightest Cluster Galaxies (BCG hereafter) themselves. These peculiar galaxies
often have very weak and extended halos that are very difficult to detect and
separate from the intrastructure diffuse light. When classical
flux measurement tools such as Sextractor are applied to the CFHTLS data, part of 
this contribution might sometimes be missed and is not included in the galaxy CFHTLS magnitude measurements. 
We may therefore underestimate the luminosity of the BCGs.
This strongly depends on the considered BCG (or BGG) and on the history of its parent
structure, but \citet{Martizzi+14}, for instance, estimated the
BCG-related Intra Cluster Light (ICL hereafter) contribution to values higher than 20$\%$ of the total BCG
luminosity. All this contribution is not missed when magnitudes are measured in the CFHTLS survey, but even when only 20$\%$ are missed, this may disfavor the FG probability estimate because the value of the
BCG magnitude is overestimated. In order to test this effect, we reran our codes assuming this
20$\%$ missing luminosity. This resulted in diminishing the
considered two-magnitude gap by $\sim$0.2 magnitude within the PFG computation.

This 20$\%$ contribution will increase the number of potential FGs,
reaching 25 structures with a PFG value higher than 50$\%$ (15 structures without the
20$\%$ increase). However, this does not change the fact that such structures have
masses preferentially lower than 2.4$\times$10$^{14}$~M$_\odot$ (see 
figure~\ref{fig:PercMasstop}). 

As a summary, we present in Table~\ref{tab:liste} the different numbers of structures for different levels
of probability to be a FG. We also list the numbers when a 20$\%$ contribution by the dominant 
galaxy is assumed.

\section{General structure properties versus PFG value}

\subsection{PFG versus M$_{200}$}

\begin{figure}[h]
\includegraphics[width=9cm,angle=0]{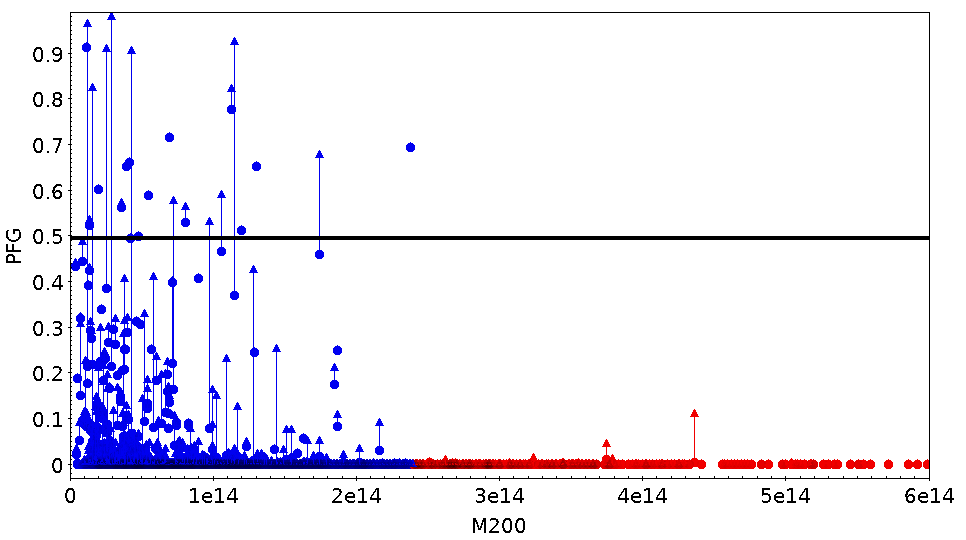}
\caption{\label{fig:PercMasstop} PFG probability for the structures to be FGs as a function of their estimated M$_{200}$ \citep{Sarron+18}. The vertical
blue arrows show the PFG increase when a 20$\%$ ICL additional contribution to the 
BCG (or BGG) magnitude is allowed. Blue dots are galaxy structures less 
massive than 2.4$\times$10$^{14}$~M$_\odot$ , and red dots are more massive than 
2.4$\times$10$^{14}$~M$_\odot$. The horizontal black line shows the 50$\%$ level for PFG.}
\end{figure}

\begin{figure}[h]
\includegraphics[width=9.5cm,angle=0]{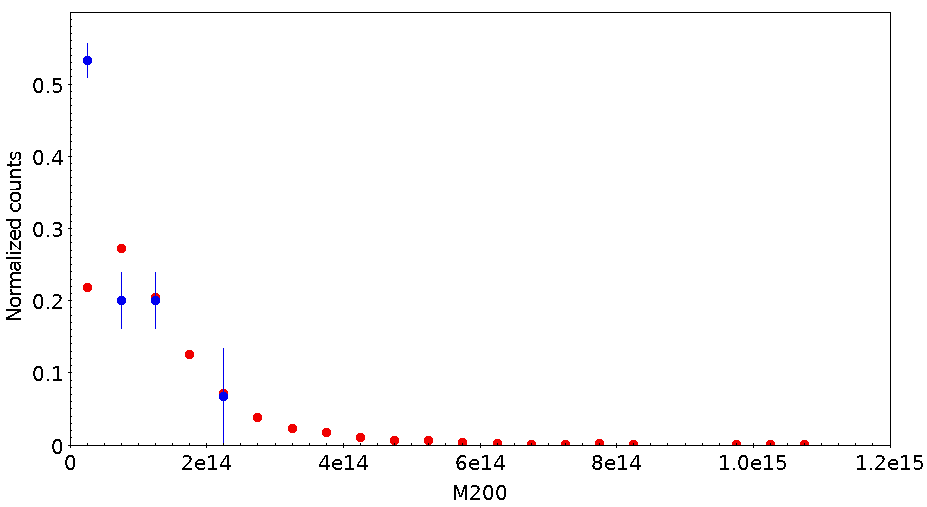}
\caption{\label{fig:mass} Arbitrarily normalized counts of the estimated
M$_{200}$ masses (in M$_\odot$ units) of the AMASCFI 
candidate structures. Red dots represent the whole sample, and blue dots show structures with a
probability 
higher than 50$\%$ to be a FG. Both samples are normalized. Error bars are
Poissonian.}
\end{figure}

We test in this subsection the general behavior of the  structures detected in the
CFHTLS Wide survey \citep{Sarron+18} in terms of variation
of their M$_{200}$ as a function of their probability PFG to be a
fossil group.
Figure~\ref{fig:PercMasstop} shows that the more massive a structure, the less
likely it is a FG, in the
sense that the PFG versus M$_{200}$ space is basically empty above
M$_{200}$$\sim$2.4$\times$10$^{14}$~M$_\odot$. Adding a 20$\%$ ICL contribution to the
BCG or BGG magnitude does not change this separation strongly.
When we limit our analysis to structures with PFG$\geq$50$\%$, we also see that these are low-mass 
structures (see Fig.~\ref{fig:mass}), mainly lower than 1.5$\times$10$^{14}$ M$_\odot$.

This value of 50$\%$ is a good compromise between the highest possible probability to be a FG
and the sample size. We detected 15 structures with such a probability, which are described in 
Table~\ref{tab:first15}.

\subsection{Spatial distribution relative to the nodes and filaments}

\begin{figure}[h]
\includegraphics[width=9.0cm,angle=0]{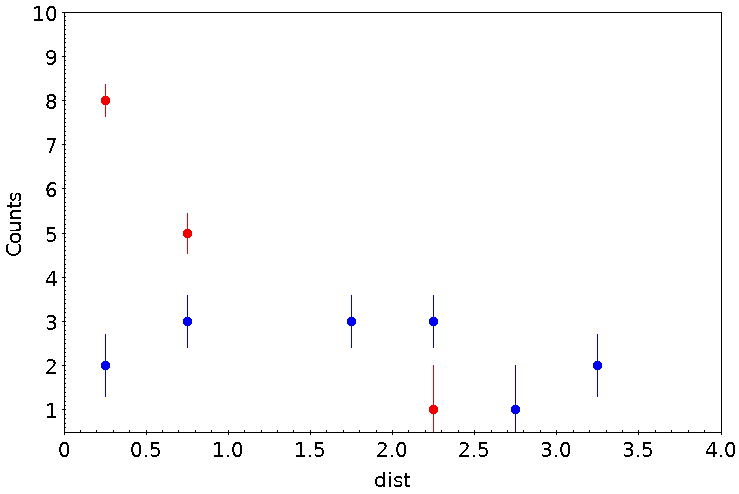}
\caption{\label{fig:distfilnode} Distributions of comoving distance (in Mpc) to the nearest filament (red dots) or node (blue dots) of the candidate FGs with PFG$\geq$50$\%$.}
\end{figure}

\begin{figure}[h]
\includegraphics[width=9cm,angle=0]{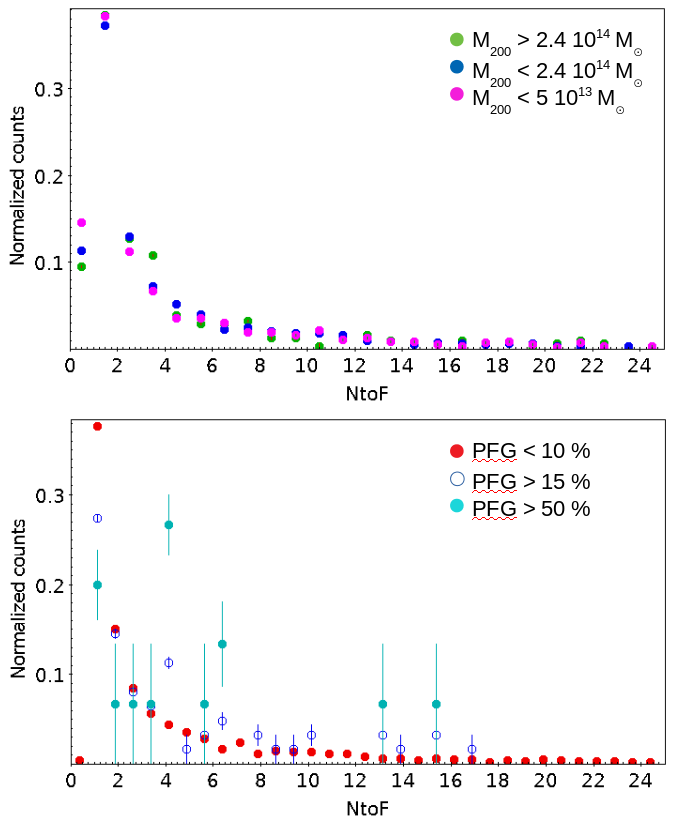}
\caption{\label{fig:MassvsPFG}Upper panel: Arbitrarily normalized counts of the
ratios NtoF between the distance to the nearest node and to the nearest filament
for different structure masses. Structures more massive than
2.4$\times$10$^{14}$ M$_\odot$ are shown in green, and structures less 
massive than 2.4$\times$10$^{14}$ M$_\odot$ are plotted in blue. Structures less massive than
5$\times$10$^{13}$ M$_\odot$ are shown in magenta.
Lower panel: Arbitrarily normalized counts of the ratios NtoF between the
distance to the nearest node and to the nearest filament for different 
PFG values. A PFG lower than 10$\%$ is plotted with filled red symbols, a  PFG higher than 15$\%$ with
open blue symbols, and a PFG higher than 50$\%$ with filled cyan symbols. Error bars are Poissonian and smaller than
the point size in the upper panel.}
\end{figure}

One of the key questions when fossil groups are studied is to know where they reside
relative to the cosmic web. They might be located in completely isolated regions,
as suggested, for example, by \citet{Adami+12}, or they might have a distribution similar
to that of other galaxy structures. The answer to these questions is a
key driver of the building mechanisms of these peculiar structures. The question is how the large magnitude gap between the first and second brightest
galaxies can be explained. They might just lack infalling galaxies, which would mean that merging events dominate infalling events. Alternatively, these infalling
galaxies might have peculiar impact parameters, which would complicate a possible capture by the structure.

Figure~\ref{fig:distfilnode} shows that the structures with the PFG values higher than
50$\%$ are preferentially closer to the filaments than to the nodes. While their distribution relative
to the nodes is more or less uniform, they exhibit a clear tendency to be located
closer than 1 Mpc to their nearest filament. More precisely, 87$\%$ are closer than 1 Mpc to the filaments, while 67$\%$ of them are farther away than 1 Mpc from
the nodes.
Moreover, we recall than the filament detection completeness is about 70$\%$, which means that structures with PFG values higher than 50$\%$ may all be
close to filaments.

We are aware that our statistics are weak (only 15 structures with PFG$\geq$50$\%$), in good agreement with the
paucity of this class of objects. Being $\geq$10 times more extended than the CFHTLS, the Kilo-Degree (KiDS) survey
\citep{deJong+15} and later the EUropean Cosmic aLl sky Investigator of the Dark universe (EUCLID) surveys 
\citep{Laureijs+11} 
will be much better adapted to detect large populations of FGs when the filaments and nodes will be
detected in these areas.

\subsection{Structure distribution within the cosmic web as a function of PFG and mass}

The previous part seems to show that structures with a PFG value higher than 50$\%$
preferentially reside close to 
the filaments and relatively far from the cosmic web nodes. We investigated whether this is
related to the mass of the considered structures.
To determine this, we split the whole structure sample into masses higher than
$2.4\times10^{14}$~M$_\odot$, 
lower than $2.4\times10^{14}$~M$_\odot$, and lower than 5$\times$10$^{13}$ M$_\odot$.
We also computed the ratio between the
distance to the nearest node and to the nearest filament (NtoF hereafter). A high value of NtoF indicates that the considered
structure is closer to its nearest filament than to its nearest node.
Figure~\ref{fig:MassvsPFG} (upper part) shows the general behavior of the NtoF
ratio. NtoF is close to 1 for the majority of the structures (as seen in the figure,
where the dominant contribution occurs slightly above NtoF$\sim$1).
Such structures are equally distant from nodes or
filaments. The figure also exhibits a tail of structures closer to filaments 
than to nodes (NtoF$\geq$1). This is an expected behavior because filaments cover a larger cosmic
volume than nodes. We see no significant effect as a function of mass: a Kolmogorov-Smirnov test shows that the probability is lower than 0.1$\%$ for the three distributions to differ.

Figure~\ref{fig:MassvsPFG} (lower part) is similar, but focuses on the PFG value effect
rather than on the mass effect. The structures that are the most 
different from the FG status (PFG lower than 10$\%$) have a similar
behavior as in Fig.~\ref{fig:MassvsPFG} (upper part), with a peak around
a NtoF$\sim$1. This means that they are not significantly more distant from the nodes than
from the filaments. Intermediate structures (PFG higher than 15$\%$) may be more distant from the nodes,
with a larger number of them exhibiting higher NtoF values. The difference 
between structures with PFG lower than 10$\%$ and PFG higher than 15$\%$ is
not very significant, however: a Kolmogorov-Smirnov test predicts a probability of only 7$\%$ 
for the two distributions to be different.
For the structures with the highest PFG value (higher than 50$\%$), the situation is
different. These structures exhibit a preferential NtoF value of $\sim$4, which is significantly
different from the abundances of the other samples at the same NtoF value. They are
statistically about four times less distant from a filament than from a node. 
Another peak is present 
around NtoF $\sim$6.5, but is only barely significant because of its large error bar.
A Kolmogorov-Smirnov test predicts a probability of 93$\%$ for the distribution with PFG
values higher than 50$\%$ to be different from the distribution with PFG values
lower than 10$\%$.
To summarize, the higher the probability of being a FG, the closer these structures appear to be to
the filaments.

\section{Conclusion}

The two main conclusions of this study have been reached by probabilistic arguments and can be
summarized as follows:

1) Only the less massive of our candidate clusters (M$_{200}\le 2.4\times
10^{14}$~M$_\odot$) can have a high  probability of being FGs.

2) Structures with the highest probabilities of being FGs are preferentially
close to their nearest cosmic filament ($\leq$1~Mpc), and their NtoF
ratio is preferentially of $\sim$4.

It is therefore tempting to say that fossil groups reside in cosmological
filaments but do not survive in cosmic nodes. 
This can be explained because in the nodes, the merging rate of structures and/or
the galaxy infalling rate are too high and are not compensated for by
the intragroup galaxy-galaxy merging rate to maintain the two-magnitude gap
observed in FGs.
In the filaments, the density (both in terms of field galaxies and other structures)
is lower than in the nodes. Fossil groups can therefore
empty their vicinity, and their mass evolution is eventually stopped by a
lack of accretable material. 
Another explanation would be that the relative velocities of the group-infalling galaxies
are high enough in the filaments to prevent galaxies from being captured
by such low-mass galaxy structures. In both cases, the galaxy-galaxy
intragroup mergers depopulate the group and create the two-magnitude gap in the
group galaxy population.

We are aware that these conclusions are based on probabilistic arguments. Only
intensive spectroscopic surveys will allow placing these results on firmer grounds.

\begin{acknowledgements}
The authors thank the referee. 
The authors thank Clotilde Laigle for useful discussions. We gratefully acknowledge financial 
support from the Centre National d'Etudes Spatiales (CNES) for many years. NM
acknowledges support by a CNES fellowship.

Based on observations obtained with XMM-Newton, an ESA science mission with
instruments 
    and contributions directly funded by ESA Member States and NASA.
    Based on observations made with ESO Telescopes 
    at the La Silla and Paranal Observatories under programmes ID 191.A-0268
    and 60.A-9302. Based on observations obtained with
    MegaPrime/MegaCam, a joint project of CFHT and CEA/IRFU, at the
    Canada-France-Hawaii Telescope (CFHT) which is operated by the
    National Research Council (NRC) of Canada, the Institut National
    des Sciences de l'Univers of the Centre National de la Recherche
    Scientifique (CNRS) of France, and the University of Hawaii. 
    This work is based in part on data products produced at Terapix
    available at the Canadian Astronomy Data Centre as part of the
    Canada-France-Hawaii Telescope Legacy Survey, a collaborative
    project of NRC and CNRS. This research has made use of the VizieR catalogue
    access tool, CDS, Strasbourg, France. This research has also made
    use of the NASA/IPAC Extragalactic Database (NED) which is
    operated by the Jet Propulsion Laboratory, California Institute of
    Technology, under contract with the National Aeronautics and Space
    Administration. This research has made use of the ASPIC database, operated 
    at CeSAM/LAM, Marseille, France. This paper uses data from the VIMOS Public
Extragalactic Redshift Survey (VIPERS). 
    VIPERS has been performed using the ESO Very Large Telescope, under the "Large
Programme" 182.A-0886. The 
    participating institutions and funding agencies are listed at
http://vipers.inaf.it. Based on data obtained with 
    the European Southern Observatory Very Large Telescope, Paranal, Chile, under
Large Program 185.A-0791, and made 
    available by the VUDS team at the CESAM data center, Laboratoire d'Astrophysique
de Marseille, France. This 
    research uses data from the VIMOS VLT Deep Survey, obtained from the VVDS
database operated by Cesam, Laboratoire 
    d'Astrophysique de Marseille, France. GAMA is a joint European-Australasian
project based around a spectroscopic 
    campaign using the Anglo-Australian Telescope. The GAMA input catalogue is based
on data taken from the Sloan Digital 
    Sky Survey and the UKIRT Infrared Deep Sky Survey. Complementary imaging of the
GAMA regions is being obtained by a 
    number of independent survey programmes including GALEX MIS, VST KiDS, VISTA
VIKING, WISE, Herschel-ATLAS, GMRT and 
    ASKAP providing UV to radio coverage. GAMA is funded by the STFC (UK), the ARC
(Australia), the AAO, and the participating 
    institutions. The GAMA website is http://www.gama-survey.org/.  This paper uses
data from the XXL survey, an international project based around an XMM Very
Large Programme surveying two 25 deg2 extragalactic fields at a depth of
$~6\times10^{-15}$ erg cm$^{2}$ s$^{-1}$ in the [0.5-2] keV band for point-like
sources. The XXL website is http://irfu.cea.fr/xxl. Multi-band information and
spectroscopic follow-up of the X-ray sources are obtained through a number of
survey programmes, summarised at http://xxlmultiwave.pbworks.com/.

\end{acknowledgements}

\bibliographystyle{aa} 
\bibliography{biblio} 

\clearpage

\appendix

\section{List of the 15 galaxy structures with PFG$\geq$50$\%$}                                                                                      

\begin{table*}[t!]
\caption{\label{tab:first15}List of the 15 structures from \citet{Sarron+18} with PFG$\geq$50$\%$ with (1) the 
corresponding CFHTLS field,  (2-3): RA and
uncertainty, (4-5): DEC and uncertainty, (6) detection signal-to-noise ratio (S/N) within the CFHTLS structure sample, (7)
redshift, (8) M$_{200}$, (9) R$_{200}$, (10) distance to the nearest node, (11) distance to the nearest filament, (12) 
probability of being a FG. 
The structure at coordinates (219.40, 56.55) is probably in a region where filament detection has been inoperative.}
\begin{tabular}{cccccccccccc}
\hline
CFHTLS field & RA&RA uncertainty&DEC&DEC uncertainty&S/N& redshift&M$_{200}$&R$_{200}$&dnode&dfil&PFG \\
&(deg)&(deg)&(deg)&(deg)& & & M$_{\odot}$& Mpc & Mpc&Mpc& $\%$ \\
\hline
\hline
W1&34.72&0.05&-9.27&0.05&6&0.22&4.20E13&0.78&1.77&0.51&50 \\
W1&36.76&0.03&-9.27&0.03&4&0.32&1.20E14&1.15&3.42&0.64&51 \\
W1&35.89&0.05&-8.69&0.05&6&0.26&1.39E13&0.56&2.59&0.17&52 \\
W1&32.20&0.05&-6.60&0.05&3&0.20&3.97E13&0.77&2.10&0.50&65 \\
W1&30.80&0.01&-9.80&0.01&4&0.66&1.30E14&1.35&0.83&0.19&65 \\
W1&36.82&0.02&-4.55&0.02&10&0.28&2.38E14&1.47&0.81&0.21&69 \\
W1&38.18&0.03&-9.77&0.03&3&0.36&6.92E13&1.00&2.07&0.32&71 \\
W1&38.69&0.03&-5.64&0.03&6&0.30&1.13E14&1.15&1.69&0.26&78 \\
W1&37.14&0.03&-5.84&0.04&4&0.30&1.15E13&0.55&0.51&0.57&91 \\
\hline
W2&136.19&0.02&-4.65&0.02&6&0.68&8.06E13&1.17&0.39&0.10&53 \\
W2&135.46&0.04&-4.33&0.04&4&0.22&1.97E13&0.61&2.41&0.19&60 \\
\hline
W3&213.46&0.01&54.78&0.01&6&0.56&4.75E13&0.94&1.92&0.80&50 \\
W3&212.57&0.06&57.69&0.03&6&0.36&3.63E13&0.77&3.15&2.31&56 \\
W3&219.40&0.07&56.55&0.04&5&0.26&4.18E13&0.79&35.54&35.64&66 \\
\hline
W4&334.64&0.02&-0.76&0.02&7&0.36&5.48E13&0.93&0.16&0.09&59 \\
\hline
\end{tabular}
\end{table*}

\end{document}